\documentclass[conference]{IEEEtran}
\IEEEoverridecommandlockouts
\usepackage{url}
\usepackage{graphicx}
\usepackage{bbm}
\usepackage{forest}
\usepackage{xcolor}
\usepackage{amsmath}
\usepackage{varwidth}
\usepackage{subcaption}
\usepackage{tikz}
\usepackage{pgf-umlsd}
\usetikzlibrary{shapes,arrows,shadows} 
\usepackage{bm}
\usepackage{amssymb}
\usepackage{algorithm}
\usepackage{algpseudocode}
\usepackage{pgf}
\usetikzlibrary{arrows,automata}
\usepackage{pgf-umlsd}
\usetikzlibrary{calc}
\usepackage[latin1]{inputenc}
\usetikzlibrary{decorations.pathreplacing}
\usetikzlibrary{shapes.geometric}

\usepackage{subcaption}

\usepackage[left=0.65in,right=0.65in,top=0.7in,bottom=1in]{geometry}

\newcommand{\hwvar}{v}
\definecolor{TUMBlau}{RGB}{0,101,189} 
\definecolor{TUMBlauDunkel}{RGB}{0,82,147} 
\definecolor{TUMBlauHell}{RGB}{152,198,234} 
\definecolor{TUMBlauMittel}{RGB}{100,160,200} 

\definecolor{TUMElfenbein}{RGB}{218,215,203} 
\definecolor{TUMGruen}{RGB}{162,173,0} 
\definecolor{TUMOrange}{RGB}{227,114,34} 
\definecolor{TUMGrau}{gray}{0.6} 

\usepackage{booktabs} 
\usepackage[normalem]{ulem} 

\normalsize

\begin{document}
%


\title{Boost Your CotS IEEE 802.15.4 Network \\ with Inter-Slot Interference Cancellation \\ for Industrial IoT }
\author{{H. Murat G\"ursu, Hansini Vijayaraghavan, Wolfgang Kellerer}\\
{Chair of Communication Networks,
Technical University of Munich,
Germany} \\
Email: \{murat.guersu, hansini.vijayaraghavan, wolfgang.kellerer\}@tum.de  }
%

%


\maketitle





\begin{abstract}

The current cellular standardization for 5G is working towards wireless advances to enable further productivity for industrial automation. However, this development will take several years. Meanwhile, the capabilities of the currently available standards should be pushed to their limits. To this end, in this work, we present results from the first inter-slot successive interference cancellation testbed using commercial off the shelf IEEE 802.15.4 sensors. Through our implementation, we have measured a throughput of $0.72$ packets per slot which doubles the currently used contention-based access, Slotted ALOHA, with a limit of $0.36$ packets per slot. The hardware effects of the boards, which degrade the successive interference cancellation performance from the theoretical limit of $1$ packet per slot, are modeled and validated through measurements. We also propose a model that can be used to calculate the expected successive interference cancellation throughput with the specific hardware available in a factory. Furthermore, our proposed model should replace the perfect physical layer assumptions for researchers to design new MAC algorithms taking practical limitations into account.
 \end{abstract}


\section{Introduction}
\label{sec:intro}

 The integration of automation in factories is gaining momentum and is limited by the capabilities of the currently deployed wireless networks. This defines the Industrial Internet of Things (IIoT) challenge. 3GPP is addressing this challenge in the release 16 of 5G \cite{3gpp_iiot}. The standardization introduced by 5G will cover use-cases that extend at least to the next 10 years. However, the products supporting the release 16 can reach the market earliest in late 2020, as the current functional freeze of standardization is scheduled for March of 2020 \cite{3gpp_release}. Another {challenge} is the unknown price of these devices and the required time to market to reach a reasonable price for all business sizes. 

\par This motivates us to investigate a solution improving the currently deployed networks for IIoT. IEEE 802.15.4 \cite{TSCH} is one of the frequently available standards in factories thanks to its energy efficiency. The PHY layer is fixed in most of the devices as this is embedded in the radio chip but the MAC layer behavior is controlled through a protocol. Thus, a lot of work in IEEE 802.15.4 \cite{wang2010dynamic} \cite{6987334} is focusing on optimizing the MAC protocol that can be updated easily for the deployed sensors. A large group of researchers focus on the coordination of the periodical monitoring of sensors to optimize the use of the wireless medium. However, periodic monitoring is not fit for IIoT, as those devices are reacting to sporadic events. Random access, investigated as grant-free in 5G networks \cite{R1-1808304}, is a flexible solution for this problem. Random access may have varying resource efficiency. {However, it has been shown that successive interference cancellation based random access algorithms can achieve a throughput matching that of the scheduled access without giving up the flexibility of random access \cite{6325214}.} Previous works on random access that consider low-latency and reliability constraints \cite{murat_acdc} \cite{polling_abbas} \cite{stefanovic2017frameless} assume a perfect physical layer performance that does not represent the practical limitations. 

\subsection{Related Work}
The hardware effect should be considered to obtain practical limits, and this is investigated in terms of radio irregularity. In \cite{zhou2006models}, the main causes of radio irregularity are summarized as irregularities with antenna angle and battery power. A recent work \cite{trenkamp2011wireless} demonstrates that sensors cannot reach the documented maximum transmission powers in their data-sheets. In another work \cite{zamalloa2007analysis}, the authors analyze hardware performance variance and calculate the packet reception rate without interference cancellation. These works neglected the impact of other hardware effects such as that of the oscillator on the SNR.

Successive interference cancellation (SIC) has attracted interest in terms of intra-slot interference cancellation (IaS-IC). An implementation of a receiver employing SIC for IEEE 802.15.4 for IaS-IC is introduced in \cite{halperin2008taking}. The packet of the strongest user, in terms of signal power, in a collision is decoded and canceled from the same time-slot. Authors in \cite{lv2011scheduling} implemented the IaS-IC for scheduling multiple packets to the same time-slot in IEEE 802.15.4. There have been many implementations of SIC that modify the physical layer such as \cite{zigzag} and \cite{mzig}. Such modifications are out of scope for us, as we intend for the improvement to work with standard-compliant IEEE 802.15.4 PHY layer. 

Combining the hardware effects and the SIC is also investigated in the state of the art. A Signal to Interference plus Noise Ratio (SINR) model, based on residual interference power for imperfect SIC, is introduced in \cite{andrews}. The authors provide bounds on the transmission capacity for imperfect successive interference cancellation wherein a fraction of the interference power is left behind after cancellation. In our work, we decompose this model into specific errors related to channel estimation and phase estimation to provide insights on the limitation of the CoTS hardware, validated through measurements.

\subsection{Contribution}
In our work, we deploy an inter-slot successive interference cancellation (IeS-IC) algorithm, that fulfills low-latency and reliability constraints with an IEEE 802.15.4 network. We measure the practical performance in terms of throughput in packets per slot and show that our implementation doubles the throughput of {the commonly deployed} Slotted ALOHA based network i.e., from $0.36$ to $0.72$. We provide an analytical throughput model, using a detailed SINR evaluation, to extend our measurements to any scenario with different hardware characteristics. We validate our model with the measurements and discuss performance with different hardware. Consequently, our model provides insights for not only the state of the current devices but also any future hardware that is foreseen to use SIC.
 
\subsection{Organization}
The rest of the work is organized as follows: Section \ref{sec:phy} introduces the SINR model and extends it with the hardware effects. Section \ref{sec:mac} introduces the MAC algorithm and the throughput model. The measurement setup is described and the results are evaluated in Sec.~\ref{sec:meas}. Finally the contributions are summarized and future directions are described in Sec.~\ref{sec:conc}.

\section{PHY Model}
\label{sec:phy}
In this section we introduce how the successive interference cancellation is implemented.
\subsection{Scenario}
In this work a cell-based star topology using a time-division multiple access system is considered. The time is divided into time-slots and only a single frequency channel is used. Multiple users are attached to a central station. The users are not transmitting regularly. They are only reacting to events. The users can receive the broadcast downlink channel, but have to contend for the resources in the uplink channel. Thus, the uplink resources are allocated in a distributed manner, using a random access algorithm, guided by the downlink feedback of the central station. The throughput of the uplink resource usage is defined as number of packets transmitted per time-slot, abbreviated as packets per slot.

Considering a typical random access algorithm like Slotted ALOHA, as the load increases, the throughput decreases quickly. Thus Slotted ALOHA is unsuitable for traffic bursts. Therefore, our implementation uses successive interference cancellation, where users transmit multiple replicas of their packets in different time-slots. If multiple users transmit in the same time-slot then this is called a collision. If a replica of a packet involved in the collision is available in a different time-slot without interference, then the interference free packet is used to reduce the interference in the collision slot. However, the channel effects have to be recreated to properly cancel it from the collision slot. For this purpose, the distortions to the signal due to the channel as well as the hardware have to be estimated. 

\subsection{{Proposed SSINR} Model}
Mobility is not considered and the channel is assumed to be time-invariant and static. The wireless channel attenuates the signal with respect to objects and the distance. Multipath propagation is assumed to exist where the copies of the signal travelling in different paths interfere at the receiver causing inter-symbol interference (ISI). 


Other than wireless effects we also investigate hardware effects such as phase noise by introducing a new parameter, $\hwvar$. 

Let us assume the interference-free received signal is
\begin{center}
	$r_i = x_i\hwvar_ih_i + n_i $,
\end{center}
where, $x_i$ is the modulated signal, $\hwvar_i$ is the coefficient representing the effect of uncertainty coming from the transmitter hardware noise, $h_i$ is the channel gain and $n_i$ is the Additive White Gaussian Noise in the channel.

The channel estimation includes constant phase drift stemming from the hardware effects\footnote{The phase noise is considered separately and not included in the channel}. 
However, the channel estimation depends on the hardware and is much worse when estimated in a collision slot. This error is defined as the mean channel estimation error and is represented as a constant $\epsilon$, as introduced in \cite{cheon2002effect}.  

For the sake of simplicity let us consider an example for user $i$. The channel of the user $i$ is estimated to be ${h_i}$ but the realization is $h_i^o$, with an error $\epsilon$,
\begin{center}
	$h_i^o = h_i(1+\epsilon)$.
\end{center}
 Hence, the residual signal, when the re-created signal with the estimated channel is removed from the received signal, $r_i$ is,
\begin{equation}
	r_i -x_ih_i = \underbrace{{x_i\hwvar_ih_i}-x_ih_i }_{\text{Hardware Noise Error}} + \underbrace{{x_ih_i\hwvar_i}\epsilon}_{\text{Channel Est. Error}} + \underbrace{n_i}_{Noise}.
\end{equation}

The SINR in a collision slot is the signal power $\gamma_i$ divided by the residual power caused by hardware noise that is  $\gamma^p_i$, the residual power caused by channel estimation error $\gamma^c_i$ and the noise power $\gamma^n_i$
\begin{equation}
	SINR_i = \frac{\gamma_i}{\gamma^p_i +\gamma^c_i+ \gamma^n_i}.
\end{equation} 


For the case with multiple rounds of successive interference cancellation, the SIC-SINR (SSNIR) of the packet for the $i^{\text{th}}$ user when $k$ packets are cancelled is given by,
\begin{equation}
	SSINR(i,\mathcal{S}, \mathcal{C}) = \Theta_i^{\mathcal{S}, \mathcal{C}} = \frac{\gamma_i}{\sum_{k}^{\mathcal{S}}\left(\gamma^p_k +\gamma^c_k + \gamma^n_k \right) + \sum_{j}^{\mathcal{S}/\mathcal{C}} \gamma_j  }
	\label{eq:ssnr}
\end{equation}
where $\mathcal{S}$ is the set of all packets transmitted at the same time and $\mathcal{C}$ represents the set of canceled packets including $i$.

\subsection{Measurements for SSINR model}\label{meas_ssinr_model}

The model parameters are obtained by performing {dedicated} measurements for each Zolertia Z1 mote transmitting to a USRP B200-mini receiver. Channel gain $h$ is estimated at the receiver. The statistics of noise, $n$, are collected from the portion of the received signal where no data packets are transmitted. The statistics of signal affected by the transmitter hardware noise, channel gain and channel AWGN, $x\hwvar h + n$, for different users are obtained from the slots where the users' packets do not face any interference. Since the noise $n$ is additive, knowing its mean and variance, the statistics of $x\hwvar h$ are calculated treating them as the combination of 2 independent random variables with different means and variances. The signal $x$ has a unit envelope. Therefore, removing the effect of the constant channel gain $h$, the statistics of $\hwvar$ are obtained.

In our work, we adopt most of the PHY receiver chain of \cite{halperin2008taking} and adapt it for inter-slot interference cancellation adding cross slot channel estimation with phase drift estimation. The phase drift of a packet at a particular time-slot can be determined by calculating the mean of the difference in phase between the ideal signal and the received signal over the length of the packet. The ideal signal is known from the interference-free replica of the packet. 

\begin{figure}[t!]
	\centering
	\includegraphics[width = 0.4\textwidth]{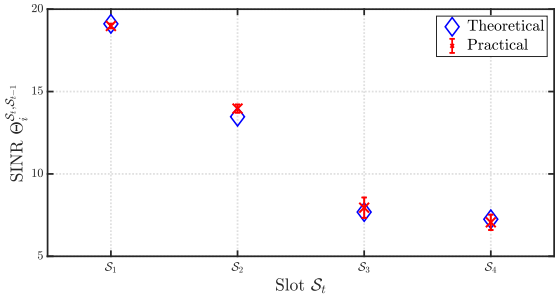}
	\caption{SIC-SINR model versus measurements showing validity of the selected parameters.}
	\label{fig:SNR_SIC}	
\end{figure}

The validity of the SIC-SINR model and the selected parameters is illustrated in Fig.~\ref{fig:SNR_SIC} where measurements are given with 95$\%$ confidence intervals and compared to Eq.~\eqref{eq:ssnr}. {The $\hwvar$ parameter is obtained for each device through dedicated measurements as mentioned in Section \ref{meas_ssinr_model}.} The typical values for the mean and variance of the $v$ parameter as observed in the measurements for a Zolertia Z1 mote are around 1 and 0.01, respectively. The $\epsilon$ value is set as $0.2$ representing the channel estimation error when $i \neq k$ and set as $0.001$ when $i = k$.
\subsection{SNR to BER mapping}

A channel model that represents the measurement setup \cite{zheng2011link} exists in reality. We will use the Rician SNR to BER mapping function $f_B( \gamma)$ for QPSK as given in \cite{simon1998unified},

\begin{align}
f_B( \gamma)& =  \frac{1}{\pi} \int_{0}^{3\pi/4} \mathcal{M}_\gamma \left( -\frac{sin^2(\pi/4)}{sin^2(\theta)} \right)	d\theta \nonumber \\ \text{      where,   }  \mathcal{M}_\gamma \left( s \right) &= \frac{1+K}{1+K - s\gamma}e^{\frac{K\gamma}{(1+K) -s\gamma}}.	
\end{align}
We set $K=4$, fitting the measurements, representing the relative strength of line of sight signal compared to non line of sight signal. The function outputs the bit error rate that can be converted to packet error rate or the decoding probability $P_d$ using the packet length $P_l$ in bits,


\begin{equation}
	P_d( \Theta_i^{\mathcal{S}, \mathcal{C}})  = \left(1-f_B\left( \Theta_i^{\mathcal{S}, \mathcal{C}}\right) \right)^{P_l}.
\end{equation}


\section{MAC Model}
\label{sec:mac}
\subsection{Query Tree Algorithm with SIC (SICQTA)}

For the MAC algorithm, we focus on an instant-feedback based random access algorithm that is the successive interference cancellation query tree algorithm (SICQTA) \cite{gursu2019hard}. This algorithm is one of two algorithms that have deterministic latency bounds along with the access codes introduced in \cite{boyd2018interference}. 


In SICQTA, every device has a unique id composed of $u$ bits. This limits the total number of devices attached to the central station to $N = 2^u$. Users transmit packets after they are queried. Queries include a part of the address bits. The initial query is an empty query and all active users $M$ answer to the query. A single bit is appended to the query after each collision, starting from the left-most bit. As each device has a unique id, this guarantees that with $u$ collisions a query is the unique address of a device.



\renewcommand{\arraystretch}{0.5}
\begin{figure}[t]
	\centering
	\begin{subfigure}[t]{0.4\textwidth}
		\centering
	\begin{tabular}{|c|c|c|c|c|c|}
	\rule{0.8cm}{0pt}&\rule{0.8cm}{0pt}&\rule{0.8cm}{0pt}&\rule{0.8cm}{0pt}&\rule{0.8cm}{0pt}&\rule{0.8cm}{0pt}\\[-2.0ex]
			\hline
		\scriptsize$\mathcal{S}_6$& \scriptsize{$\mathcal{S}_3$} & \scriptsize{$\mathcal{S}_2$} & 
		\scriptsize{$\mathcal{S}_i$} & \scriptsize{$\mathcal{S}_5$} &  \scriptsize{$\mathcal{S}_4$}  \\				\hline
			\scriptsize{A,B,C,D} & \scriptsize{A,B} & \scriptsize{A,B} & 
		\scriptsize{A} & \scriptsize{C,D} &  \scriptsize{C}  \\	
		\hline
	\end{tabular}
	\begin{tabular}{cc}
	\rule{1cm}{0pt}&\rule{1cm}{0pt}\\[-3.3ex]
\end{tabular}
	\begin{forest}
		for tree={circle,draw,minimum size=0.5cm,l=1cm,s sep=1cm}
		[\scriptsize{A,B,C,D}
		[\scriptsize{A,B},edge label={node[midway,left,font=\scriptsize]{}}
		[\scriptsize{A,B},edge label={node[midway,left,font=\scriptsize]{}}
		[\scriptsize A,edge label={node[midway,left,font=\scriptsize]{}}]
		[\scriptsize B,dotted,edge label={node[midway,left,font=\scriptsize]{}}]
		] [\scriptsize{},dotted,edge label={node[midway,left,font=\scriptsize]{}}
		]			]
		[\scriptsize{C,D},dotted,edge label={node[midway,left,font=\scriptsize]{}}
		[\scriptsize{C,D},edge label={node[midway,left,font=\scriptsize]{}}
		[\scriptsize C,edge label={node[midway,left,font=\scriptsize]{}}]
		[\scriptsize D,dotted,edge label={node[midway,left,font=\scriptsize]{}}]
		]				[\scriptsize{},dotted,edge label={node[midway,left,font=\scriptsize]{}}
		] ]]		;
	\end{forest}
\end{subfigure}
	\tiny
	\caption{SICQTA worst case with $M=4$}
	\label{fig:wc_sicqta}
\end{figure}
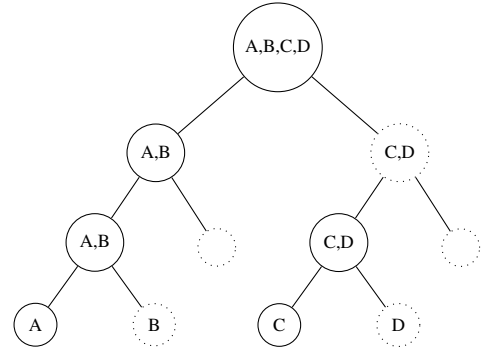


A detailed example for the SICQTA is given in Fig.~\ref{fig:wc_sicqta} for $M=4$ and $u=3$, users A,B,C,D have addresses $\{000,001,100,101\}$ respectively. In the first slot, all the devices are queried and it is a collision. On the second and third slot, $0xx$ and $00x$ are queried, respectively. Both are collisions. In the following slot, $000$ is queried and it is a success. $001$ is not queried, as the central station recovered the packet B from slot $\mathcal{S}_2$ and $\mathcal{S}_1$. $001$ and $01x$ are not queried and skipped. Thus, $10x$ is queried, that results in a collision. Consequently, $100$ is queried and is a success. The gateway recovered D from slot $\mathcal{S}_5$ and $\mathcal{S}_4$. Finally, the algorithm is terminated.

\subsection{User Activity Analysis} 
\label{sec:chan_est}   

We are interested in the practical throughput performance of the complete system. Some of the throughput loss is caused by the MAC algorithm. Thus, we first introduce the MAC throughput analysis without the PHY considerations.

In our setting, the users have a $u=3$ bit address. Hence, there are $N=8$ users.

\renewcommand{\arraystretch}{0.5}
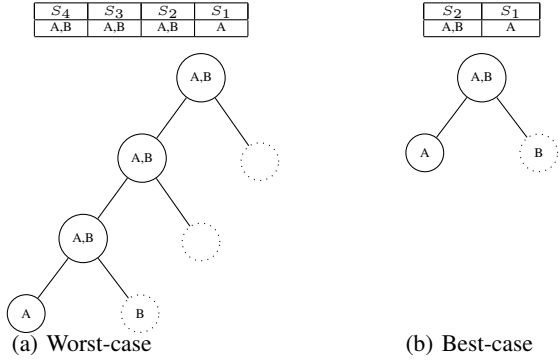
\begin{figure}[t]
	\centering
\hfill
	\begin{minipage}[t]{0.40\linewidth}
	\centering
	\begin{center}
	\begin{tabular}{|c|c|c|c|}
	\rule{0.23cm}{0pt}&\rule{0.23cm}{0pt}&\rule{0.23cm}{0pt}&\rule{0.23cm}{0pt}\\[-1.4ex] 
	\hline
	\tiny{$S_4$} & \tiny{$S_3$} & \tiny{$S_2$}  & \tiny{$S_1$} \\
	\hline
	\tiny{A,B} & \tiny{A,B} & \tiny{A,B}  & \tiny{A}  \\				\hline
\end{tabular}
	\end{center}
	\begin{forest}
		for tree={circle,draw,minimum size=0.5cm,l=0.5cm,s sep=1cm}
		[\tiny{A,B}
		[\tiny{A,B},edge label={node[midway,left,font=\scriptsize]{}}
		[\tiny{A,B},edge label={node[midway,left,font=\scriptsize]{}}
		[\tiny{A},edge label={node[midway,left,font=\scriptsize]{}}]
		[\tiny {B},dotted,edge label={node[midway,left,font=\scriptsize]{}}]
		]		[\tiny {},dotted,edge label={node[midway,left,font=\scriptsize]{}}]
		]		[\tiny {},dotted,edge label={node[midway,left,font=\scriptsize]{}}]]		;
	\end{forest}
\end{minipage}
\hfill
\begin{minipage}[t]{0.40\linewidth}
	\centering
		\begin{center}
	\begin{tabular}{|c|c|}
\rule{0.35cm}{0pt}&\rule{0.35cm}{0pt}\\[-1.4ex]
\hline
	\tiny{$S_2$} & \tiny{$S_1$}  \\
		\hline
		\tiny{A,B} &  \tiny{A}  \\				\hline
	\end{tabular}
\end{center}
	\begin{forest}
		for tree={circle,draw,minimum size=0.5cm,l=0.5cm,s sep=1cm}
		[\tiny{A,B}
		[\tiny{A},edge label={node[midway,left,font=\scriptsize]{}}]	
		[\tiny {B},dotted,edge label={node[midway,left,font=\scriptsize]{}} 
		]]		;
	\end{forest}
\end{minipage}
\hfill
\begin{subfigure}{.2\textwidth}
	\caption{Worst-case}
	\label{fig:2211}
\end{subfigure}
\hfill
\begin{subfigure}{.2\textwidth}
	\caption{Best-case}
\label{fig:21}
\end{subfigure}
\hfill
\caption{Possible tree structures for $M=2$ active user and $u=3$ maximum depth of the tree.}\label{fig:confs}
\end{figure}

%
%

Consider the scenario that $M=2$ users have collided with the first query. Two possible ways in which this tree could split after this stage is as shown in Fig. \ref{fig:confs}. If a collision of two users happens multiple times, as in Fig.~\ref{fig:2211}, two slots are wasted. The users are selecting slots uniformly. The probability of occurrence of the worst-case configuration given $M=2$  is $\frac{1}{2^2} \cdot \frac{1}{2^2}$, as two users have done the same selection twice and is $\frac{1}{2^1}$ for the best-case. The throughput is calculated as the number of slots used versus number of packets received, e.g, $\frac{2}{4}$ for the worst-case and $\frac{2}{2}$ for the best-case scenario. In this way the throughput of all scenarios can be calculated. Thus, not all loss comes from the physical layer but also due to lack of coordination in MAC.

The average MAC throughput $\rho$ is calculated by the formula,

\begin{center}
	$\rho = \sum_{o} P_{\text{occ}_o} \cdot \rho_o$\\
\end{center}
where $P_{\text{occ}_o}$ is the probability of occurrence of the $o^\text{th}$ scenario and $\rho_o$ is the throughput of that scenario. A scenario is the way a tree is formed as illustrated in Fig.~\ref{fig:wc_sicqta} and Fig.~\ref{fig:confs}.



 
\subsection{Scenario Resolution Probability}
 
In this subsection we take PHY imperfections into consideration, i.e., we take decoding errors into account. The probability of resolution taking decoding errors into account for scenario $o$, is a new parameter $P_{\text{res}_o}$. The throughput can be calculated as,
 \begin{equation}
 	\rho = \sum_o P_{\text{occ}_o} \cdot P_{\text{res}_o} \cdot \rho_o.
 	\label{eq:tpt}
 \end{equation}
The resolution probability $P_{\text{res}_o}$ is assumed 1 for MAC throughput. 
 

The probability of resolution can be calculated as the joint probability distribution for successfully decoding all slots as in,
\begin{equation}
P_{\text{res}_o} = \prod_{t=1}^n P_d{(\Theta_i^{\mathcal{S}_o^t, \mathcal{S}_o^{t-1}} )},
\label{eq:resi_td}
\end{equation}
where $\mathcal{S}_o^t$ is the list of all users that transmitted in slot $t$ for scenario $o$. Slots in the frame can be depicted as $[\mathcal{S}_o^1, \mathcal{S}_o^2, \cdots, \mathcal{S}_o^n]$ where $n$ is the number of slots in a frame and $\mathcal{S}_o^0$ is defined as an empty set.

 $i$ is a user, whose packet can be decoded from set $\mathcal{S}_o^t$ given that packets of all users in set $\mathcal{S}_o^{t-1}$ are cancelled from $\mathcal{S}_o^t$. The $\Theta_i^{\mathcal{S}_o^t,\mathcal{S}_o^{t-1}}$ depicts the SINR for packet of user $i$ in slot $t$ after the set $\mathcal{S}_o^{t-1}$ is cancelled. And example of how sets are defined can be seen in Figures~\ref{fig:wc_sicqta} and \ref{fig:confs}. In both examples $\mathcal{S}_1$ contains only A, while $\mathcal{S}_2$ contains A and B. The slots are ordered following the tree structure.

%
%
%
 Idle slots are left out of the list as they have no effect on the resolution. 
However, if two sets are equal, it means that the same packets are transmitted on two different slots. Thus, a success is the decoding of either one of these slots. The probability of that can be calculated by checking the failure of either of the values. For instance, given a scenario where the initial slot is repeated $k$ times, Eq.~\eqref{eq:resi_td} can be re-organized as,
\begin{figure}[t!]
	\centering
	\includegraphics[width = 0.4\textwidth]{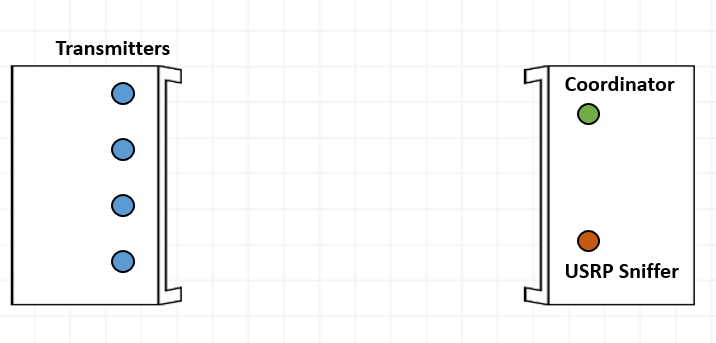}
	\caption{Experimental setup}\label{fig:mac_setup}
	\vspace{-3mm}
\end{figure}

\begin{equation}
P_{\text{res}_o} = \left(1-\left(1- P_d(\Theta_i^{\mathcal{S}_o^1} ) \right)^k\right) \cdots P_d(\Theta_i^{\mathcal{S}_o^n,\mathcal{S}_o^{n-1}} ) .
\label{eq:t_div}
\end{equation}
The hidden assumption here is, the realization on two different time-slots is not exactly the same in terms of noise and if cancellation is not possible with one, it may be possible with the other slot. This modification demonstrates that repetitions for successive interference cancellation increase the reliability exponentially. This will be further evaluated in the results section.


\section{Measurements}
\label{sec:meas}

\subsection{Experimental setup}
 A testbed of 5 Zolertia Z1 nodes and a USRP B200-mini receiver is set up as shown in Figure \ref{fig:mac_setup}.

Four Z1 nodes, depicted in blue, act as the transmitters and each has a direct line of sight path to a fifth Z1 node, depicted in green, acting as the network coordinator and central station. All of the Z1 used are 5 years old representing a mid-life of an industrial sensor. The USRP, depicted in red, is set up to act as a sniffer. It has a direct line of sight path to the transmitters and receives all the packets. All nodes in the network communicate on the IEEE 802.15.4 channel 26 which is centered at 2480 MHz. A single slot is of duration $4$ ms and the firmware of the sensors is implemented with OpenWSN.



We are measuring a SICQTA scenario of $M=4$ active users with address size of $u=5$ {as most inter-slot SIC algorithms result in scenarios where up to 4 users are cancelled.} The users are set to transmit $128$ byte packets. We repeated this scenario for $100$ consecutive frames. The sniffed data\footnote{The measurement data and scripts are available at \url{https://gitlab.lrz.de/lkn_measurements/sicqta_measurements} } captured with the SDR is processed in MATLAB.

\subsection{Processing measurements}
The raw data file sampled at 4 MHz is read into MATLAB and the beginning of a frame is detected by comparing the signal strength to a threshold by manually observing the interference and noise signal levels at the receiver.

Firstly, all time-slots with a single packet are decoded. The location of the replicas can be traced back thanks to the tree structure. Then the replicas are canceled from time-slots with collisions. After each SIC iteration, a canceled slot is decoded and the decoded packets are again canceled from other slots. A scenario is considered successful if all packets involved in the scenario are canceled. Following this, the measurement success probability is calculated. 

\subsection{Evaluations}

\begin{table}[t!]
	\begin{center}
		\begin{tabular}{|l |c |c|c|}
			\hline 
			\textbf{Scenario} & \textbf{$P_{\text{res}_o}=1$} & \textbf{Measured}& \textbf{Model}\\
			\hline  
			\hline 	 
			4 users & 0.875 & 0.5837 & 0.6026\\
			\hline 
			3 users & 0.8344 & 0.6926 & 0.6465\\
			\hline 
			2 users & 0.7917 & 0.7273 & 0.7495\\
			\hline 
		\end{tabular}
		\caption{Comparison of theoretical and Practical throughputs}\label{tab:thru}
	\end{center}
\end{table}

\begin{table}[t!]
	\begin{center}
		\begin{tabular}{|l |c |c |}
			\hline 
			\textbf{Scenario} & \textbf{Meas.} & \textbf{Model}\\
			\hline  
			\hline 	 
			2221 & 1 & 0.9997\\
			\hline 
			221 & 0.98 & 0.9954\\
			\hline
			21 & 0.9 & 0.9324\\
			\hline
			\hline 	 
			3321 & 0.9& 0.8757\\
			\hline
			3221 & 0.82& 0.7498\\
			\hline
			321 & 0.74& 0.7024\\
			\hline
			\hline
			3311 & 0.98& 0.9391\\
			\hline
			3121 & 0.9& 0.9324\\
			\hline 
			311 & 0.87& 0.7532\\ 
			\hline
		\end{tabular}
		\hspace{0.2cm}
		\begin{tabular}{|l |c |c |}
			\hline 
			\textbf{Scenario} & \textbf{Meas.} & \textbf{Model}\\
			\hline  
			\hline 	 
			44211 & 0.84& 0.8643\\
			\hline \hline
			4321 & 0.59& 0.5125\\
			\hline \hline
			4311 & 0.56& 0.5496\\
			\hline \hline
			422121 & 0.88& 0.9282\\
			\hline
			42121 & 0.81& 0.8695\\
			\hline
			42211 & 0.7& 0.7263\\
			\hline
			4211 & 0.64& 0.6804\\
			\hline \hline
			4111 & 0.66& 0.7297\\
			\hline
		\end{tabular}
		\caption{Resolution probability $P_{\text{res}_o}$ for the different scenarios of 2, 3 and 4 user collisions}\label{tab:pres2}
	\end{center}
\end{table}



 The practical throughput for different number of active users in packets per slot are summarized in Tab.~\ref{tab:thru} through measurements and the model given with Eq.~\ref{eq:resi_td}. In Tab.~\ref{tab:pres2} we have detailed the throughput results in packets per slot for each scenario.

Firstly, the measurement results confirm that IeS-IC can be used to improve throughput for contention based access using the CotS IEEE 802.15.4 hardware with throughput reaching at least $0.58$ and up to $0.72$. {This doubles the maximum throughput of the commonly deployed Slotted Aloha scheme which is 0.368 \cite{kleinrock1975packet}.} In terms of latency, when $M=4$ the radio latency is at maximum $72$ ms representing 3 frames and on average $24$ ms. In terms of data-rate, using contention based access and a single channel, this translates to a reliable $170$ kbit/s. In case all 16 channels are used, this can be further boosted approximately to a reliable $2.7$ Mbit/s. The results are promising for industrial scenarios with sporadic activity and a delay constraint of $100$ ms. 

Taking a deeper look at the results, the $P_{\text{res}_o}=1$ throughput is $0.79$ packets per slot, that is $21\%$ lower than the maximum MAC throughput of 1 packet per slot. The measurements, where $P_{\text{res}_o}$ is replaced with measured values, demonstrate that there is a further $7\%$ throughput loss due to PHY problems with 2 users. The results demonstrate that the MAC algorithm can be further optimized before PHY layer issues are resolved when $M=2$ users are considered.

As SICQTA performs better with more users e.g., $M=4$, the $P_{\text{res}_o}=1$ throughput is $12.5\%$ lower than the maximum of 1 packet per slot. As demonstrated with the measurements, $0.3$ packets per slot is lost due to PHY failure. Higher throughput is only possible with excessive use of SIC. Thus, we see that the higher throughput we expect, the importance of PHY effects increases. The SSINR model becomes significantly important when we want to reach maximum throughput of 1 packet per slot.

Furthermore, the model shows a good match overall. Interference cancellation up to 4 users is quite common in most of the inter-slot SIC algorithms such as IRSA or Frameless ALOHA. The tree algorithms, due to their structure, enforce a higher order of cancellation. For all these algorithms, the perfect PHY assumption can be replaced with the model we provide here for realistic evaluations. Another approach is clearly the use of better hardware that can decrease the worst-case latency if no re-transmissions are needed. 



%

\begin{figure}[t!]
	\centering
	\includegraphics[width = 0.5\textwidth]{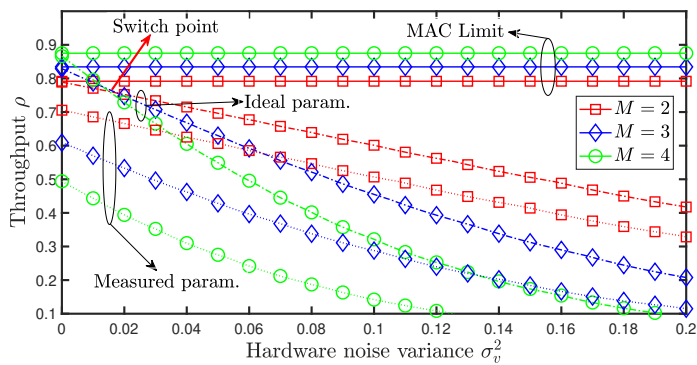}
	\caption{Sweep analysis for the hardware noise variance $\sigma_v^2$ vs throughout as calculated with Eq.~\ref{eq:tpt}.Two different set of system parameters, $\{\epsilon,n,\sigma_v^2\}$ are evaluated as ideal $\{\epsilon=0,n=0,\sigma_v^2=0\}$ and measured  $\{\epsilon=0.2,n=0.1,\sigma_v^2=0.001\}$ variables.} \label{fig:sims}
\end{figure}

Encouraged by the match of the model and the measurements we do a sweep analysis with the model. In Fig.~\ref{fig:sims} we have plotted MAC throughput varying the hardware noise variance. The results show that even with a really high channel gain and no estimation error, after $\sigma_\hwvar^2=0.03$, having $M=2,3,4$ outputs the same throughput as the collisions with more users have a increased decoding error probability. The results show that hardware effects pose a big limitation to implement inter-slot IC. In case new IEEE 802.15.4 hardware will be bought for SIC, the hardware noise variance can be evaluated to decide if it lies before the switch point as depicted in Fig.~\ref{fig:sims}. In case the hardware noise variance of the device lies on the right hand side of the switch point, that hardware cannot benefit fully from SIC.

\begin{figure}[t!]
	\centering
	\includegraphics[width = 0.5\textwidth]{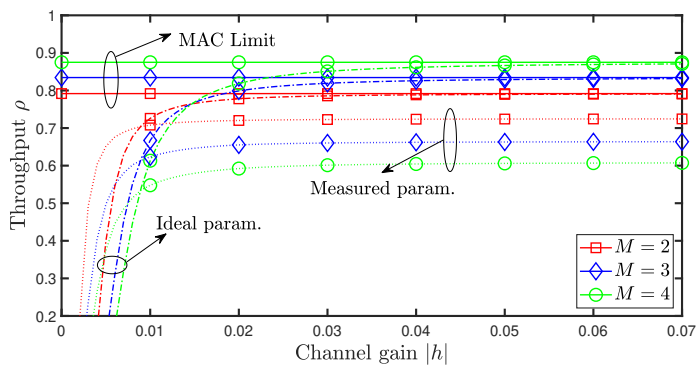}
	\caption{Sweep analysis for channel gain $|h|$ vs throughput as calculated with Eq.~\ref{eq:tpt}.} \label{fig:sims_h}
\end{figure}

The effect of the channel gain is illustrated in Fig.~\ref{fig:sims_h} with a similar sweep analysis. The measured parameters reflect that after a channel gain of $0.01$, the throughput does not increase due to estimation error and hardware noise. With ideal parameters the channel gain required to reach the MAC limit is around $0.04$. This evaluation reflects the required transmission power, antenna selection and placement of the sensor to reach the required MAC throughput for a given hardware setting. 

\section{Conclusion}
\label{sec:conc}

In this work, we have demonstrated a wireless communication solution for supporting industrial internet of things paradigm. To this extend, we have shown that CotS IEEE 802.15.4 chips, in a contention based access, can achieve a similar throughput compared to scheduled access, solving the problem of event-based access of sensors. This is achieved through using a successive interference cancellation MAC algorithm in the receiver that does inter-slot interference cancellation. To the best of our knowledge, it is the first experimental evaluation of inter-slot IC for IEEE 802.15.4, which is particularly challenging due to signal variations from slot to slot due to cheap hardware.

We introduced a practical SIC-SINR model that we have validated through experimental results in our testbed. This model can be used to design MAC algorithms that are aware of these limitations or to evaluate other SIC-MAC algorithms. 

Finally, we have once more demonstrated through measurements that the hardware effects are limiting the high throughput promised by MAC algorithms. Further measurements should be extended to better hardware to conclude what the maximum achievable performance with SIC for CoTS IEEE 802.15.4 sensors is and how MAC algorithms can deal with this limitation.

\bibliographystyle{IEEEtran}
\bibliography{hw_est}
%

\end{document}